\documentclass[%
 reprint,
superscriptaddress,
preprintnumbers,
nofootinbib,
 amsmath,amssymb,
 aps,
 prl,
floatfix,
]{revtex4-2}

\usepackage{enumitem}

\usepackage{graphicx}%
\usepackage{dcolumn}%
\usepackage{bm}%

\usepackage{epsfig}
\usepackage{amsmath}
\input{epsf}
\usepackage{psfrag}
\usepackage{xcolor}
\usepackage{pdfpages}
\usepackage[normalem]{ulem}
\usepackage{tikz}
\usepackage{booktabs,cellspace}
\usepackage{graphicx}
\usepackage{subcaption} 
\usepackage[labelformat=simple]{subfig}

\makeatletter
\AtBeginDocument{\let\LS@rot\@undefined}
\makeatother
\newcommand{\bea}{\begin{eqnarray}}
\newcommand{\eea}{\end{eqnarray}}

\newcommand{\nn}{\nonumber}

\def\z{\zeta}
\def\cE{\mathcal{E}}
\def\cD{\mathcal{D}}
\def\DGLAP{\text{DGLAP}}
\def\EEC{\mathrm{EEC}}
\newcommand{\GEEC}[2]{\mathrm{E}^{#1}\mathrm{E}^{#2}\mathrm{C}}
\begin{document}

\preprint{MIT-CTP 5894}
\preprint{CALT-TH 2025-023}
\preprint{CPTNP-2025-025}

\title{Quantum Scaling in Energy Correlators Beyond the Confinement Transition}

\author{Cyuan-Han Chang}
\email{cchang10@uchicago.edu}
\affiliation{Leinweber Institute for Theoretical Physics, University of Chicago, Chicago, Illinois 60637, USA}

\author{Hao Chen}
\email{hao\_chen@mit.edu}
\affiliation{Center for Theoretical Physics - a Leinweber Institute, Massachusetts Institute of Technology, 77 Massachusetts Avenue, Cambridge, MA 02139, USA}

\author{Xiaohui Liu}
 \email{xiliu@bnu.edu.cn}
 \affiliation{Center of Advanced Quantum Studies, School of Physics and Astronomy, Beijing Normal University, Beijing, 100875, China}
 \affiliation{Key Laboratory of Multi-scale Spin Physics, Ministry of Education, Beijing Normal University, Beijing 100875, China}

\author{David Simmons-Duffin}
\email{dsd@caltech.edu}
\affiliation{Walter Burke Institute for Theoretical Physics, Caltech, Pasadena, California 91125, USA}

\author{Feng Yuan}
\email{fyuan@lbl.gov}
\affiliation{Nuclear Science Division, Lawrence Berkeley National
Laboratory, Berkeley, CA 94720, USA} 

\author{Hua Xing Zhu}%
 \email{zhuhx@pku.edu.cn}
\affiliation{School of Physics, Peking University, Beijing 100871, China}%
\affiliation{Center for High Energy Physics, Peking University, Beijing 100871, China}

\begin{abstract}
We study the QCD scaling behavior of the small-angle Energy-Energy Correlator (EEC), focusing on the transition between its perturbative pre-confinement and non-perturbative post-confinement regimes. Applying the light-ray Operator Product Expansion (OPE), we develop a formalism that describes the scaling of the EEC with the input energy $Q$ in the transition and the post-confinement region, where the latter quantum scaling is determined by the $J=5$ DGLAP anomalous dimension. A key result of our work is a novel connection between the light-ray OPE and the dihadron fragmentation function (DFF), where we show that the non-perturbative OPE coefficients correspond to moments of the DFF. This finding establishes a new paradigm for studying hadronization. Our theoretical predictions are validated against Monte Carlo simulations for both $e^+e^-$ and $pp$ collisions, showing excellent agreement. The potential role of the quantum scaling in the precision determination of $\alpha_s$ is also discussed.
\end{abstract}

\maketitle

\section{Introduction}

Since its discovery in the 1970s, Quantum Chromodynamics~(QCD) has been the established theory of the strong interactions, governing a rich array of phenomena from the confinement of quarks and gluons to the emergence of jets and the formation of quark-gluon plasma~\cite{Gross:2022hyw}. 
Despite five decades of progress, the complex, non-perturbative nature of QCD continues to pose significant challenges, necessitating the development of novel theoretical and experimental tools. 

\begin{figure*}[htbp]
  \centering

  \begin{subfigure}[b]{0.31\linewidth}
    \includegraphics[width=\linewidth]{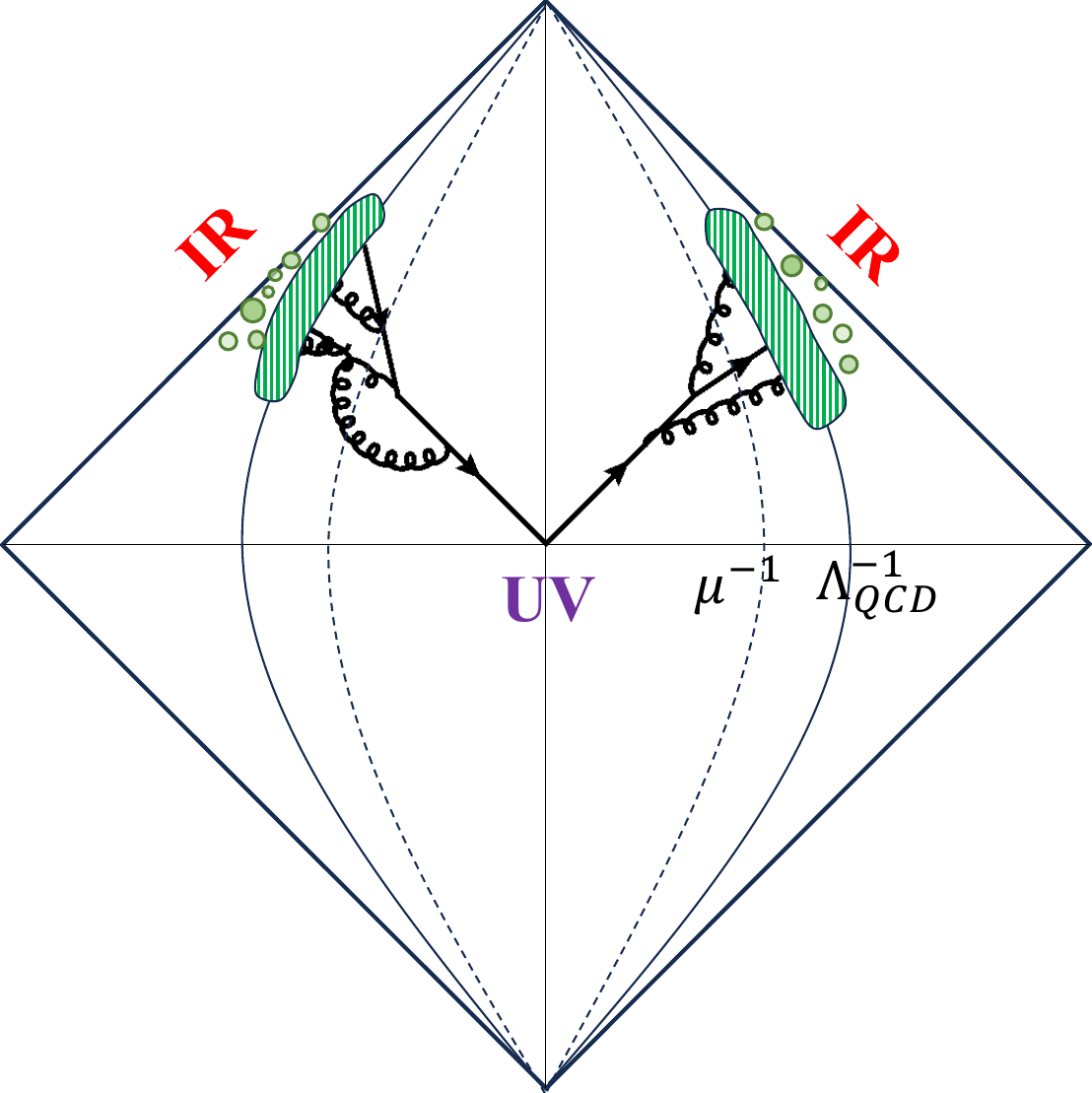}
    \caption{}
    \label{fg:illustration}
  \end{subfigure}
  \hspace{1cm}
  \qquad
  \begin{subfigure}[b]{0.4\linewidth}
    \includegraphics[width= \linewidth]{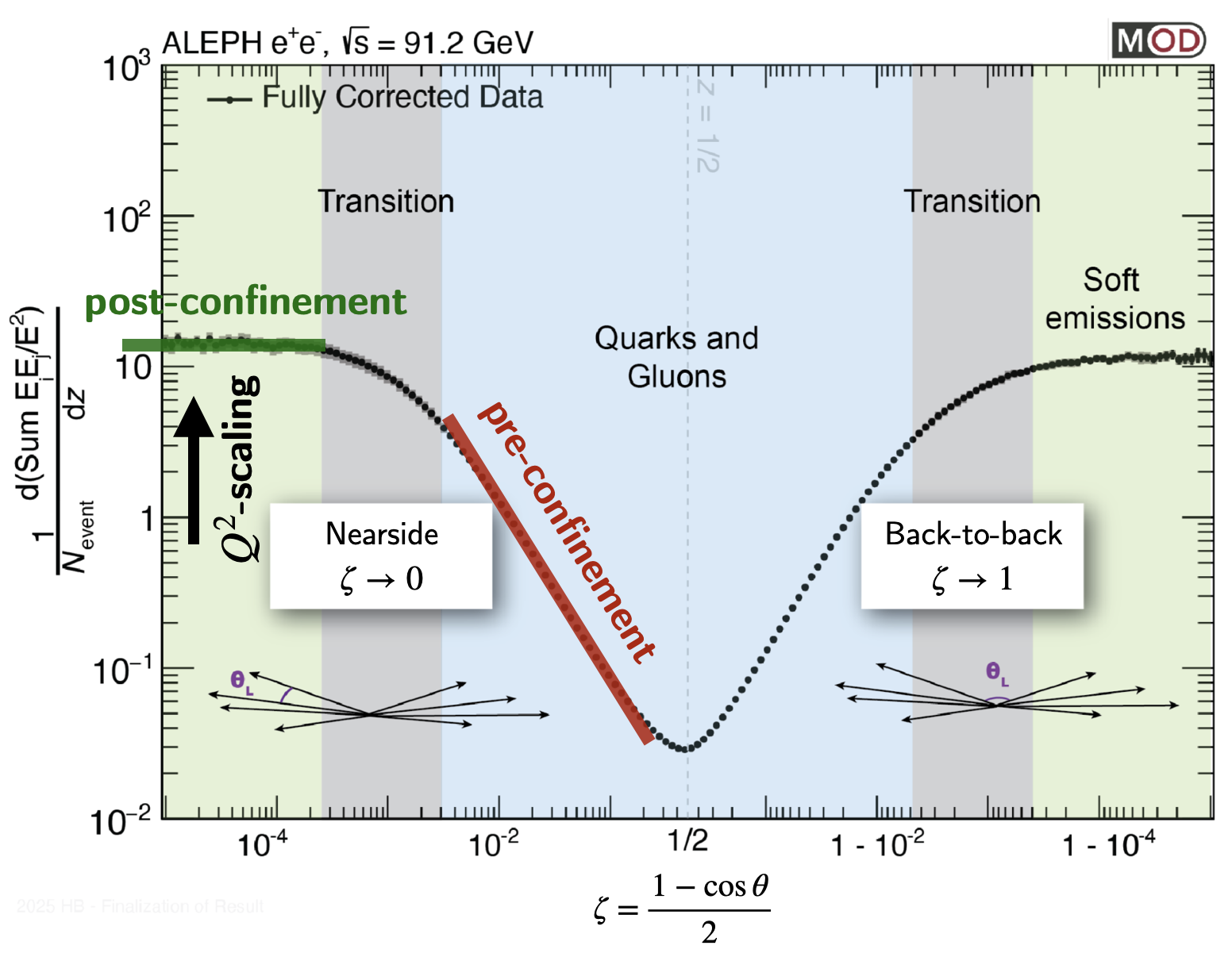}
    \caption{}
    \label{fg:LEP_scaling}
  \end{subfigure}

  \caption{(a) A schematic illustration of quark and gluon hadronizing into color-neutral hadrons; (b) Illustration for different scaling behavior of EEC. Fig.~\ref{fg:LEP_scaling} is adapted from \cite{Bossi:2025xsi}.}
  \label{fig:both-figures}
\end{figure*}

A key phenomenon in QCD that remains a mystery is the process of hadronization~\cite{Field:1977fa}, a term describing the real-time dynamical transition of quarks and gluons into color-neutral hadrons, see Fig.~\ref{fg:illustration}. There are two popular approaches to describe this process: phenomenological models implemented in parton showers, such as string and cluster hadronization~\cite{Andersson:1983ia,Webber:1983if}; and factorization theorems, which formally separate perturbative and non-perturbative dynamics via universal fragmentation functions~\cite{Collins:1981uw}. To progress beyond the current paradigm, new experimental probes sensitive to the formation of hadrons are needed, along with theoretical advances that can interpret them.

Recently, it has been revealed from CMS Open Data analysis~\cite{Komiske:2022enw} and direct experimental measurements~\cite{Tamis:2023guc,CMS:2024mlf,ALICE:2024dfl,STAR:2025jut,CMS:2025ydi,ALICE:2025igw}, that the Energy-Energy Correlator~(EEC)~\cite{Basham:1978bw} exhibits two scaling regimes when measured in jet substructure, which we call ``pre-confinement scaling" and the ``post-confinement plateau," along with a smooth transition between them, as illustrated in Fig.~\ref{fg:LEP_scaling}. The pre-confinement scaling regime has been understood in terms of perturbative quark/gluon splitting~\cite{Konishi:1979cb,Dixon:2019uzg}, along with systematic methods for incorporating non-perturbative corrections~\cite{Lee:2024esz,Chen:2024nyc}. The post-confinement plateau is a new feature that has not been fully understood and has inspired many recent studies~\cite{Liu:2024lxy,Barata:2024wsu,Chicherin:2023gxt,Csaki:2024zig}.

The EEC is a correlation function of the energy operator ${\cal E}(\boldsymbol{n})$ in a state created from $e^+e^-$ annihilation at center-of-mass energy $Q$, denoted as $\langle {\cal E}(\boldsymbol{n}_1) {\cal E}(\boldsymbol{n}_2) \rangle_Q$. The energy operator is the simplest example of a light-ray operator~\cite{Hofman:2008ar,Kravchuk:2018htv}, and it can be expressed as an integral of the local energy-momentum tensor~\cite{Sveshnikov:1995vi,Hofman:2008ar,Kravchuk:2018htv},
\begin{equation}
{\cal E}(\boldsymbol{n}) = \lim_{r \to \infty} r^2 \int_0^\infty dt\,  T_{0 \boldsymbol{n}}(t,\vec{r}) \,.
\label{eq:definitionofE}
\end{equation}
Taking the limit, (\ref{eq:definitionofE}) can be viewed as an integral along future null infinity, at a location $\boldsymbol{n}$ on the celestial sphere.
The action of ${\cal E}(\boldsymbol{n})$ on a free particle state in the direction $\boldsymbol{n}$ gives its energy
$ {\cal E}(\boldsymbol{n}) | \boldsymbol{p} \rangle = \omega_{\boldsymbol{p}} \delta^{(2)}\left(\boldsymbol{n} - \frac{\boldsymbol{p}}{| \boldsymbol{p}|} \right) | \boldsymbol{p} \rangle$.
The energy operator has obvious importance in perturbative QCD (pQCD), due to its infrared and collinear safety properties~\cite {Sterman:1975xv}, and has also drawn significant interest from a theoretical point of view; see~\cite{Moult:2025nhu} for a recent review.

In this paper, we present two complementary perspectives on the $Q^2$-scaling of the EEC in the post-confinement plateau, providing insight into the dynamical hadronization transition.
Our first perspective is based on the light-ray Operator Product Expansion (OPE)~\cite{Kologlu:2019mfz,Chang:2020qpj}. Originally formulated in conformal field theory, the light-ray OPE has recently also found applications in QCD~\cite{Chen:2020adz,Chen:2023zzh,Chen:2024nyc}. It expands a product of energy operators in a series of light-ray operators, whose scaling dimensions and Lorentz spins govern the dependence of the EEC on $Q^2$ and the angle, respectively. We focus here on the $Q^2$-scaling in the post-confinement and dynamical transition regimes.

Our second, complementary, perspective provides a description in terms of fragmentation functions. We demonstrate how the $Q^2$-scaling in the post-confinement regime can be described by the dihadron fragmentation function (DFF) formalism, a framework widely applied in hadron physics~\cite{deFlorian:2003cg,Majumder:2004wh,Majumder:2004br,Jaffe:1997hf,Bianconi:1999cd,Bacchetta:2002ux,Bacchetta:2012ty,Zhou:2011ba,Cocuzza:2023vqs,Pitonyak:2023gjx}. By bridging these two views, we establish a novel connection between the light-ray OPE and the phenomenological DFF formalism, a link that merits further exploration. To conclude, we show that the predictions from our OPE framework are in agreement with Pythia simulations, validating our approach.

\section{Light-Ray OPE Analysis}

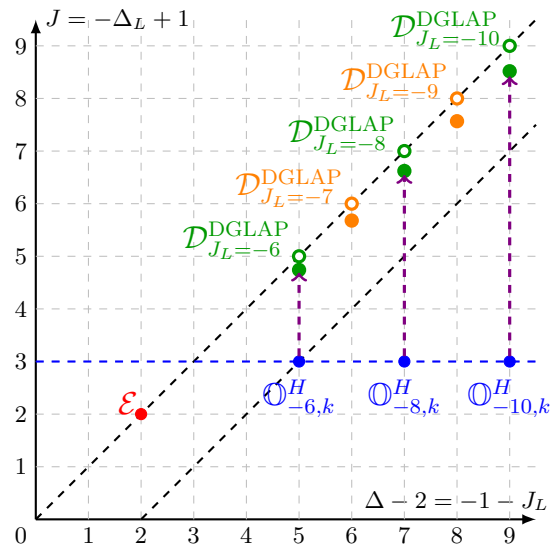
\begin{figure}[htbp]
\begin{center}
\begin{tikzpicture}[scale=0.7]
    \pgfmathsetmacro{\xmax}{9.5}
    \pgfmathsetmacro{\ymax}{9.5}
    
    \draw[black, thick, -latex] (0,0) -- (\xmax,0) node[above, xshift=-30pt] {$\Delta-2=-1-J_L$};
    \draw[black, thick, -latex] (0,0) -- (0,\ymax) node[right] {$J=-\Delta_L+1$};
    
    \foreach \x in {1,2,...,9}
        \draw[lightgray, dashed] (\x,0) -- (\x,\ymax);
    \foreach \y in {1,2,...,9}
        \draw[lightgray, dashed] (0,\y) -- (\xmax,\y);
    
    \foreach \x in {1,2,...,9}
        \node[below] at (\x,0) {\x};
    \foreach \y in {1,2,...,9}
        \node[left] at (0,\y) {\y};
    
    \node[below left] at (0,0) {0};
    
    \draw[black, thick, dashed] (0,0) -- (\xmax,\ymax);

    \draw[black, thick, dashed] (0+2,0) -- (\xmax,\ymax-2);

    \draw[blue, thick, dashed] (0,3) -- (\xmax,3);
    
    \filldraw[red] (2,2) circle (3pt);
    \node[right] at (1.4,2.2) {\large $\color{red} \mathcal{E}$};   
    
    \foreach \x in {5,7,...,9} {
        \filldraw[blue] (\x,3) circle (3pt);
        \pgfmathtruncatemacro{\label}{-1-\x} 
        \node at (\x, 2.3) {\large $\color{blue} \mathbb{O}_{\label,k}^H$};
        
	\draw[green!60!black, fill=white, very thick] (\x,\x) circle (3pt);

	\pgfmathsetmacro{\y}{\x - 0.1*ln(\x)*ln(\x)}
	\draw[green!60!black, fill=green!60!black, very thick] (\x, \y) circle (3pt);
        
        \pgfmathsetmacro{\tempy}{\x - 0.12*ln(\x)*ln(\x)}
        \draw[->, very thick, violet, dashed] (\x,3.1) -- (\x,\tempy);
        
        \pgfmathsetmacro{\tempy}{\x + 0.3}
        \node[left] at (\x, \tempy) {\large $\color{green!60!black} \mathcal{D}^{\text{DGLAP}}_{J_L=\label}$}; 
    }
    \foreach \x in {6,8} {
        \colorlet{tempcolor}{orange}
	\draw[tempcolor, fill=white, very thick] (\x,\x) circle (3pt);

	\pgfmathsetmacro{\y}{\x - 0.1*ln(\x)*ln(\x)}
	\draw[tempcolor, fill=tempcolor, very thick] (\x, \y) circle (3pt);
        
        \pgfmathsetmacro{\tempy}{\x - 0.12*ln(\x)*ln(\x)}
        
        \pgfmathtruncatemacro{\label}{-\x-1} 
        \pgfmathsetmacro{\tempy}{\x + 0.3}
        \node[left] at (\x, \tempy) {\large $\color{tempcolor} \mathcal{D}^{\text{DGLAP}}_{J_L=\label}$}; 
    }
\end{tikzpicture}
\caption{Chew-Frautschi plot illustrating the light-ray OPE for the EEC in the free hadron region and detector matching. Blue points are the locations of double twist detectors in the free hadron theory and purple arrows represent detector matching to leading-twist detectors in QCD. The detector positions in the free theory (open green circles) shift to their QCD-corrected locations at the scale $\mu$ (solid green points). Interaction between hadrons may induce odd $J_L$ (orange) contributions.}
\label{fig:CF_plot} 
\end{center}
\vspace{-5.ex}
\end{figure}

A light-ray operator $\cD(\boldsymbol{n})$ associated to a position $\boldsymbol{n}$ on the celestial sphere is characterized by a Lorentz spin $J_L \in \mathbb{C}$, which is its weight under a boost in the $\boldsymbol{n}$ direction. We also assign $\mathcal{D}(\boldsymbol{n})$ a dimension $\Delta_L$ equal to minus its mass dimension --- the minus sign is conventional because in CFT, $\cD(\boldsymbol{n})$ transforms like a primary operator at infinity \cite{Hofman:2008ar,Kravchuk:2018htv}. In a non-conformal theory, $\Delta_L$ can depend on a renormalization scale via the running coupling, $\Delta_L(J_L,\alpha_s(\mu^2))$ \cite{Caron-Huot:2022eqs,Chang:2025zib}. The energy operator ${\cal E}$ has $J_L=-3$ and $\Delta_L=-1$ (independent of scale).

The work \cite{Chang:2025zib} argues that pQCD possesses an intrinsic set of renormalized light-ray operators $\mathcal{D}_{J_L,i}(\boldsymbol{n},\mu^2)$, labeled by $J_L$, a Regge trajectory index $i$, and depending on a renormalization scale $\mu$. These light-ray operators can be visualized on a Chew-Frautschi plot, which displays $1-\Delta_{L}$ vs.\ $-1-J_L$ for each Regge trajectory, see Fig.~\ref{fig:CF_plot}. The scaling dimension $-\Delta_{L,i}(J_L,\alpha_s(Q^2))$ characterizes the growth of matrix elements $\langle\mathcal{D}_{J_L,i}(\boldsymbol{n},\mu^2)\rangle_Q$ with $Q$.

Furthermore, \cite{Chang:2025zib} argues that any measurement of hadrons at infinity can be {\it matched\/} in the regime $Q\gg \Lambda_{\rm QCD}$ onto a linear combination of detectors $\mathcal{D}_{J_L,i}(\boldsymbol{n},\mu^2)$ of pQCD. From this perspective, to determine the scaling of $\langle\cE(\boldsymbol{n}_1)\cE(\boldsymbol{n}_2)\rangle_Q$ with $Q$, we should match ${\cal E}(\boldsymbol{n}_1)\cE(\boldsymbol{n}_2)$ (viewed as an operator in the EFT of hadrons) onto a linear combination of QCD detectors. This matching is nonperturbative. However, by Lorentz-invariance, it is diagonal in boost weight: hadronic operators with definite $J_L$ match onto QCD light-ray operators with the same $J_L$.

Thus, our first task is to decompose ${\cal E}(\boldsymbol{n}_1)\cE(\boldsymbol{n}_2)$ into hadronic operators with definite boost weight $J_L$. This is accomplished with the light-ray OPE in the EFT of hadrons:
\begin{align}
\label{eq:lightrayopehadrons}
{\cal E}(\boldsymbol{n}_1)\cE(\boldsymbol{n}_2) &= \sum_{J_L,k} A_{J_L,k} C_{J_L}(\boldsymbol{n}_1,\boldsymbol{n}_2,\partial_{\boldsymbol{n}_2}) \mathbb{O}^H_{J_L,k}(\boldsymbol{n}_2).
\end{align}
Here, each hadron operator $\mathbb{O}^H_{J_L,k}$ has definite boost weight $J_L$, and a label $k$ representing different hadron operators. The differential operator 
\begin{align}
C_{J_L}(\boldsymbol{n}_1,\boldsymbol{n}_2,\partial_{\boldsymbol{n}_2}) = |\boldsymbol{n}_{12}|^{-6-J_L}(1+O(|\boldsymbol{n}_{12}|)),
\end{align}
where $\boldsymbol{n}_{12}=\boldsymbol{n}_{1}-\boldsymbol{n}_{2}$,
is entirely determined by Lorentz symmetry and sums over celestial descendants. (The operators $\mathbb{O}^H_{J_L,k}$ can also have nontrivial transverse spin \cite{Chang:2020qpj,Chen:2020adz}, but we suppress this for brevity.) We emphasize that the operators $\mathbb{O}^H_{J_L,k}$ and OPE coefficients $A_{J_l,k}$ can be defined and computed entirely in perturbation theory in the EFT of hadrons.  

In the post-confinement regime, we can approximate hadrons as nearly free massless particles. In this approximation, the OPE (\ref{eq:lightrayopehadrons}) consists of a contact term proportional to the energy-squared operator $\cE^{(2)}$, and a sum of double-twist light-ray operators $[\cE\cE]_{n,j}$ with boost weights $J_L=-6-2n-j$, transverse spin $j$, and classical dimension $\Delta_L = 2 \Delta_{L,\cE}=-2$. The precise form of these operators and their OPE coefficients will not be important for us --- only their quantum numbers. For simplicity, we will focus on spherically symmetric states, so that nonzero transverse spin operators do not contribute \cite{Chang:2020qpj,Chen:2020adz}.

\begin{table}
  \centering
  \begin{tabular}{c|c|c|c|c|c}
    \toprule
    & $\cE$  &  $\vec{\cD}^{\DGLAP}_{J_L}$ & $\cE^{(m)}$ & $\z$ & $\Lambda_{\rm QCD}$, $Q$ \\ 
    \midrule
    Lorentz spin $J_L$ & $-3$  & $J_L$ & $-m-2$ & $2$ & $0$\\
    dimension $-\Delta_L$ & $1$ & $-2-J_L$ & $m$ & $0$ & $1$ \\
    \bottomrule
  \end{tabular}
  \caption{Lorentz spin~(boost) and classical scaling dimension of
    various quantities appearing in this paper.}
  \label{tab:1}
\end{table}

In high-energy scattering $Q\gg \Lambda_{\text{QCD}}$, we can match the $\mathbb{O}^H_{J_L,k}$ onto perturbative detectors~\cite{Chang:2025zib}. As illustrated in Fig.~\ref{fig:CF_plot}, the twist-2 DGLAP detectors $\vec{\cD}^{\DGLAP}_{J_L}=(\cD^{\DGLAP}_{J_L,q},\cD^{\DGLAP}_{J_L,g})^T$ have the largest $-\Delta_L$ for a fixed $J_L$ and hence give the dominant contribution in the matching 
\begin{align}\label{eq:detector_matching}
\mathbb{O}^H_{J_L,k} \approx \vec{B}_{J_L,k}(\mu^2; \Lambda_{\text{QCD}}^2)\cdot  [\vec{\cD}^\DGLAP_{J_L}]_R(\mu^2),
\end{align}
where $\mu$ is the factorization scale, $[\vec{\cD}^\DGLAP_{J_L}]_R$ are renormalized detectors and the classical behavior of matching coefficients $\vec{B}_{J_L,k}$ is $\Lambda_{\text{QCD}}^{4+J_L}$ by dimensional analysis. Combining \eqref{eq:lightrayopehadrons} and \eqref{eq:detector_matching}, we obtain the leading light-ray OPE for $\langle\cE \cE \rangle_Q$ in the regime $\z<\Lambda_{\text{QCD}}^2/Q^2$:
\begin{align}
\label{eq:celestialblockexpansion}
\langle\cE\cE\rangle_Q &\approx \sum_{J_L} f_{J_L}(\zeta) \vec R_{J_L}(\mu^2;\Lambda_{\text{QCD}}^2)\cdot \langle  [\vec{\mathcal{D}}^{\text{DGLAP}}_{J_L}]_R(\mu^2)\rangle_Q.
\end{align}
Here, $f_{J_L}(\zeta)=\zeta^{\frac{-6-J_L}{2}}(1+O(\zeta))$ is a celestial block, obtained by acting with the differential operator $C_{J_L}(\boldsymbol{n}_1,\boldsymbol{n}_2,\partial_{\boldsymbol{n}_2})$ on a light-ray one-point function~\cite{Kologlu:2019mfz}. The coefficient $\vec R_{J_L}(\mu^2;\Lambda_{\text{QCD}}^2)$ consists of products of OPE coefficients $A_{J_L,k}$ and matching coefficients $\vec B_{J_L,k}$, summed over $k$. Finally, the matrix element $\langle [\vec{\mathcal{D}}^{\text{DGLAP}}_{J_L}]_R(\mu^2)\rangle_Q$ is evaluated in a spherically symmetric state (so it is angle-independent). The sum runs over all values of $J_L$ that appear in the OPE (\ref{eq:lightrayopehadrons}).

Let us make the free hadron approximation, so that the $J_L$ appearing in the sum are those of the double-twist operators $[\cE\cE]_{n,0}$: $J_L=-6-2n$. We furthermore focus on small angles so that we can approximate the celestial block by its leading term: $f_{J_L}(\zeta)\approx \zeta^{\frac{-6-J_L}{2}}=\zeta^n$. Making these substitutions, and defining $\EEC(\z,Q)=\langle\cE \cE \rangle_Q/Q^2$, we can write
\begin{align}\label{eq:eec_LR_OPE}
& \EEC(\zeta,Q) \approx Q^2 \sum_{n\geq 0} (\zeta Q^2)^n 
\nonumber\\
&\quad \times \vec{R}_{-6-2n}(\mu^2;\Lambda_\text{QCD}^2)\cdot
\left(
\frac{\langle  [\vec{\mathcal{D}}^{\text{DGLAP}}_{J_L=-6-2n}]_R(\mu^2)\rangle_Q}{Q^{4+2n}}
\right)
\end{align}
The quantity in parentheses has vanishing classical dimension, and hence the leading term $(n=0)$ in the OPE analysis predicts the classical $Q^2$-scaling for the height of the collinear plateau in $\EEC(\z,Q)$ ~\cite{Basham:1978bw, ALICE:2024dfl,Liu:2024lxy, Barata:2024wsu}.

This analysis can be immediately generalized to energy correlators with higher (possibly complex) energy weighting. We use $\cE^{(\alpha)}$ to denote a detector measuring hadrons with weighting $E^\alpha$, where $\mathrm{Re}(\alpha) >0$. The generalization of \eqref{eq:eec_LR_OPE} to $\GEEC{\alpha}{\beta}=\langle\cE^{(\alpha)} \cE^{(\beta)} \rangle_Q/Q^{\alpha+\beta}$ is
\begin{align}\label{eq:geec_LR_OPE}
   &\GEEC{\alpha}{\beta}(\z,Q) = Q^2 \sum_{n\geq 0} (\zeta Q^2)^{n} \nn\\
   &\;\; \times\vec{R}^{\alpha,\beta}_n(\mu^2;\Lambda_{\text{QCD}}^2) \cdot \frac{\langle  [\vec{\mathcal{D}}^{\text{DGLAP}}_{-\alpha-\beta-4-2n}]_R(\mu^2)\rangle_Q}{Q^{2+\alpha+\beta+2n}},
\end{align}
where the special case $\vec{R}_{n}^{1,1}$ reduces to $\vec{R}_{-6-2n}$ in \eqref{eq:eec_LR_OPE}. 

Following~\cite{Chen:2024nyc}, we can further study the quantum scaling violation based on \eqref{eq:geec_LR_OPE}. To see the quantum scaling behavior, we strip off the classical $Q$-dependence
\begin{align}
    F_{\alpha,\beta}(\zeta Q^2, Q^2) \equiv \frac{1}{Q^2}\GEEC{\alpha}{\beta} (\zeta,Q).
\end{align}
The scaling behavior is controlled by the RG equation for $[\vec{\cD}^\DGLAP_{J_L}]_R$
\begin{align}\label{eq:Drun}
\frac{d}{d\ln \mu^2} [\vec{\cD}^\DGLAP_{J_L}]_R = \widehat{\gamma}_{\scriptscriptstyle -J_L-1} \cdot [\vec{\cD}^\DGLAP_{J_L}]_R,
\end{align}
where $\widehat{\gamma}_J=\frac{\alpha_s}{4\pi} \widehat{\gamma}^{(0)}_J+\mathcal{O}(\alpha_s^2)$ is the time-like DGLAP anomalous dimension matrix~\cite{Dixon:2019uzg}. 
It coincides with the space-like anomalous dimension at leading order~\cite{Gribov:1972ri}, following from the reciprocity relation~\cite{Basso:2006nk,Caron-Huot:2022eqs,Chen:2020uvt}. As a consequence, we conclude that spin $J=5$ controls the quantum scaling of the post-confinement plateau. We note that in the pre-confinement regime the reciprocity relation also works for the EEC in a fixed-coupling approximation~\cite{Kologlu:2019mfz,Dixon:2019uzg,Korchemsky:2019nzm}.
 
By solving the RG equation, we introduce an evolution operator $U_{J_L}(\mu_1^2,\mu_2^2)$ to resum large logarithms related to the scaling violation
\begin{align}\label{eq:resum_result}
    &F_{\alpha,\beta}(\zeta Q^2, Q^2) =\sum_{n\geq 0} (\zeta Q^2)^n \vec{R}^{\alpha,\beta}_n(\mu_J^2,\Lambda_{\text{QCD}}^2) \cdot \nn\\
 &  U_{J_L}(\mu_J^2,\mu_H^2) \cdot \frac{\langle  [\vec{\mathcal{D}}^{\text{DGLAP}}_{J_L}]_R(\mu_H^2)\rangle_Q}{Q^{-J_L-2}}\Big|_{J_L=-\alpha-\beta-4-2n},
\end{align}
where the hard and jet scales are chosen to be $\mu_H \sim \mathcal{O}(Q)$, $\mu_J\sim \mathcal{O}(\Lambda_{\text{QCD}})$.
This is main prediction for the EEC in the post-confinement and transition regions from the light-ray OPE perspective.
Below, we will compare Eq.~\eqref{eq:resum_result} with a Monte Carlo simulation at leading logarithmic accuracy. Under such an approximation, 
the evolution kernel $U_{J_L}(\mu_1^2,\mu_2^2)$ is
\begin{align}\label{eq:LL_kernel}
    U^{\text{LL}}_{J_L}(\mu_1^2,\mu_2^2) = \left[\frac{\alpha_s(\mu_2^2)}{\alpha_s(\mu_1^2)}\right]^{\widehat{\gamma}^{(0)}_{-J_L-1}/\beta_0}\,,
\end{align}
where $\beta_0=\frac{11}{3}C_A - \frac{4}{3}n_f T_F$ is the one-loop beta function constant in QCD. 

To summarize, the spectrum of operators $\mathbb{O}^H_{J_L,k}$ in the hadron light-ray OPE (\ref{eq:lightrayopehadrons}) determines the celestial blocks in the small angle expansion of the EEC (\ref{eq:celestialblockexpansion}). The coefficient of each celestial block $f_{J_L}(\zeta)$ then evolves with scale according to the anomalous dimension of the pQCD detector with largest $-\Delta_L$ at the given $J_L$. These statements continue to hold in the presence of hadron interactions (which can modify the spectrum of $J_L$'s appearing). We give some examples in the supplementary materials.

\begin{figure}[htbp]
    \includegraphics[width=\linewidth]{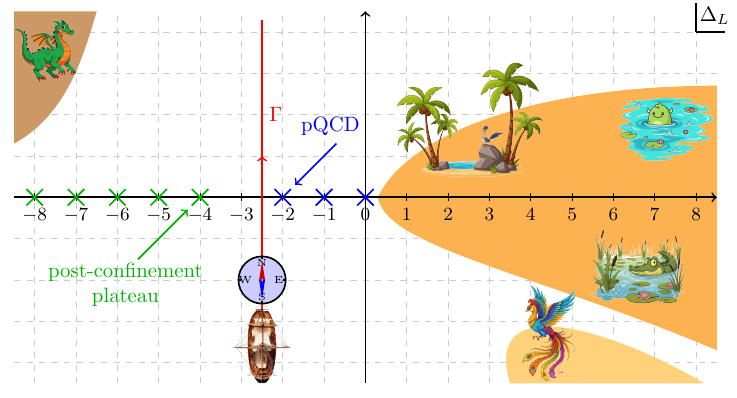}
    \caption{A conjectured contour in $\Delta_L$ plane for light-ray OPE of EEC in the QCD high-energy collision. Little knowledge is known about complex structure in $\Delta_L$.}
    \label{fg:contour_plot}
\end{figure}

The above analysis gives an expansion for the EEC in the $\zeta Q^2 \ll \Lambda_{\text{QCD}}^2$ limit. However, in the pre-confinement regime, where $\zeta Q^2 \gg \Lambda_{\text{QCD}}^2$, the expansion breaks down, and one should instead use the light-ray OPE in pQCD. It is interesting to ask whether there is a way to connect these two different expansions. Here, we conjecture a formula for the EEC for generic $\zeta Q^2/\Lambda_{\text{QCD}}^2$ which can reproduce two different expansions in the pre-confinement and post-confinement regimes.

The main idea is to start with one of the expansions, say Eq.~\eqref{eq:eec_LR_OPE}, and rewrite it in a form that converges even for generic $\zeta Q^2/\Lambda_{\text{QCD}}^2$. In \eqref{eq:eec_LR_OPE}, we can write the sum over double-twist operators as a contour integral over $\Delta_L$, where the contour encircles $\Delta_L=-4-2n$. One can then deform the contour to a new one $\Gamma=c+i\mathbb{R}$ that runs along the imaginary direction for some real number $c$. (This is reminiscent of the Sommerfeld-Watson transform in Regge theory, see e.g.\ \cite{Costa:2012cb}, where one turns a sum over spins into a contour integral.) After the contour deformation, we obtain:
\begin{gather}
    \text{EEC}(\zeta, Q)=\sum_i  \int_{\Gamma} \frac{d \Delta_L}{2 \pi i}f_{\Delta_L-\tau_i}(\zeta)\left(\frac{\Lambda_{\rm QCD}}{Q}\right)^{\Delta_L+2} \nonumber
    \\
    \times C_{\Delta_L, i}\left(\mu^2/\Lambda^2_{\text{QCD}}\right)  \frac{\left\langle\mathcal{D}_{\Delta_L, i}(\mu^2)\right\rangle_Q}{Q^{-\Delta_L}} \,,
    \label{eq:master}
\end{gather}
where the summation is over the $i$-th Regge trajectory of light-ray operators labeled with twist $\tau = \Delta_L - J_L$. Ignoring quark masses and assuming the existence of Regge trajectories of light-ray operators for QCD, Eq.~\eqref{eq:master} is the most generic formula one can write consistent with Lorentz symmetry and dimension analysis. We conjecture that this formula describes the EEC in the $\zeta \ll 1$, $\Lambda_{\text{QCD}}^2/Q^2 \ll 1$ limit, for generic $\z Q^2/\Lambda_{\text{QCD}}^2$.

Note that we are expanding an IR measurement in terms of a set of UV detectors ${\cal D}_{\Delta_L, i}$, where the OPE coefficients $C_{\Delta_L, i}$ are in general non-perturbative, except for the pQCD pole at $\Delta_L \sim -2$. In Fig.~\ref{fg:contour_plot}, we show a conjectured contour $\Gamma$ and complex structure of the coefficient $C_{\Delta_L,i}$ for the leading twist-2 trajectory. In the post-confinement regime $\zeta Q^2/\Lambda_{\text{QCD}}^2 \ll 1$, one should close the contour to the left, and the poles at $\Delta_L=-4,-5,-6,\ldots$ reproduce the expansion Eq. \eqref{eq:eec_LR_OPE}. In the pre-confinement regime $\zeta Q^2/\Lambda_{\text{QCD}}^2 \gg 1$, we close the contour to the right. The pole at $\Delta_L\sim -2$ gives the leading term in pQCD~\cite{Chen:2021gdk,Chen:2023zzh}, and the pole at $\Delta_L\sim -1$ reproduces the leading power corrections \cite{Chen:2024nyc}. Away from these poles, very little is known about the general analytic structure in the complex $\Delta_L$ plane.

\section{Relation to the dihadron fragmentation}
The non-perturbative nature of the post-confinement EEC allows the light-ray OPE analysis to be framed in terms of dihadron fragmentation, a topic that has received extensive attention in the literature~\cite{deFlorian:2003cg,Majumder:2004wh,Majumder:2004br,Jaffe:1997hf,Bianconi:1999cd,Bacchetta:2002ux,Bacchetta:2012ty,Zhou:2011ba,Cocuzza:2023vqs,Pitonyak:2023gjx}.
Specifically, we consider the scenario where in the high energy limit $Q^2 \gg \Lambda^2_{\rm QCD}$, the dominant contribution to the post-confinement EEC comes from a single parton fragmenting into a dihadron pair. In particular, the invariant mass of the hadron pair is related to their angular separation by $m^2/4 = z_1 z_2 Q^2 \zeta $, where $z_i$ is the energy fraction carried by each hadron. In this picture, the standard EEC factorization form for small $\zeta$ values~\cite{Dixon:2019uzg} remains valid, but the perturbative EEC jet function, see~\cite{Dixon:2019uzg} or the supplemental material (SM), is replaced by the non-perturbative mass-dependent DFF, see, e.g., ~\cite{Pitonyak:2023gjx}. %
The all order factorization reads
\begin{align}
& \text{EEC}(\zeta, Q)= \int\limits_0 \limits^1 \!d x \, x^2 \vec{H}(x, \frac{Q^2}{\mu^2}) \int \limits_0 \limits^\infty dm^2 \delta\left( \zeta - \frac{m^2}{4 x^2 \xi_1 \xi_2 Q^2} \right) \nonumber\\
& \cdot \sum_{(a, b)} \int d \xi_1 d \xi_2 \frac{\xi_1 \xi_2}{4} \vec{D}_{h_a h_b}\left(\xi_1, \xi_2 ; m^2 ; \mu^2 ; \Lambda_{\mathrm{QCD}}^2\right)
\label{eq:fac}\,,
\end{align}
where $\vec{H} = (2 H_q, H_g)$~\footnote{The factor of $2$ accounts for the quark and anti-quark contributions, which we assume to be the same by charge symmetry.} is the hard function that produces the fragmenting parton with momentum fraction $x$. $\vec{D}_{h_ah_b} = (D_{h_ah_b/q},D_{h_ah_b/g})$ is the mass-dependent DFF. 
Note that this Eq.~\eqref{eq:fac} reduces to the perturbative factorization in \cite{Dixon:2019uzg} when $h_a$ and $h_b$ comes from independent fragmentation (details in SM).

In the post-confinement regime, we assume $\vec{D}_{h_a h_b}$ have analytic expansion around $0< m^2 \ll \Lambda^2_{\rm QCD}$.
Taylor expansion on $m^2$ gives
\begin{equation}
\label{eq:fragexp}
\frac{{\rm EEC}(\zeta,Q) }{Q^2}  = 
 \sum_{n \ge 0} (4 Q^2\zeta)^{n} 
\vec{{\cal H}}_{4+2n}(\frac{Q^2}{\mu^2}) 
 \cdot
 \vec{{\cal J}}^{(n)}_{1,1}(\mu^2;\Lambda_{\rm QCD}^2)  \,,  
\end{equation}
where $\vec{{\cal H}}_{J-1}$ is the moment of the hard function  
$\vec{{\cal H}}_{J-1}(Q^2/\mu^2) \equiv \int_0^1 dx\, x^{J-1}  \vec{H}(x;Q^2/\mu^2) $, and the non-perturbative jet function is given by
\begin{gather}
\vec{{\cal J}}^{(n)}_{\alpha,\beta}(\mu^2;\Lambda_{\rm QCD}^2) = 
\frac{1}{n!}\sum_{(a,b)}
\int_0^1 d\xi_1 d\xi_2 \, \xi_1^{1+\alpha+n}  \xi_2^{1+\beta+n} \nonumber
\\
\partial_{m^2}^n \vec{D}_{h_ah_b}(\xi_1,\xi_2;m^2;\mu^2;\Lambda_{\rm QCD}^2) \Big|_{m^2=0} \,.
\end{gather} 
Critically, Eq.~(\ref{eq:fragexp}) is identical in form to the light-ray OPE in Eq.~(\ref{eq:eec_LR_OPE}). This allows us to identify the non-perturbative OPE coefficients $\vec{R}_{-6-2n}$ with the moment of the $n$-th derivative of the DFF 
\bea\label{eq:CandJ} 
\vec{R}_{-6-2n}(\mu^2; \Lambda_{\rm QCD}^2) = 4^n \vec{{\cal J}}^{(n)}_{1,1}(\mu^2;\Lambda_{\rm QCD}^2)  \,, 
\eea 
and the detectors $[\vec{{\cal D}}_{J_L}^{\text{DGLAP}}]_R$ with the moments of the hard function $\vec{H}$,  
\bea\label{eq:DandH} 
\langle [\vec{{\cal D}}_{J_L = -6 - 2n }^{\text{DGLAP}}]_R \rangle  
= Q^{4+2n } \vec{{\cal H}}_{4+2n}(Q^2/\mu^2) \,. 
\eea 

Eq.~\eqref{eq:CandJ} and~\eqref{eq:DandH} established a dictionary between the light-ray OPE formalism and the DFF factorization theorem. To confirm this correspondence, we check the consistency of the renormalization group evolution. The hard function $\vec{\cal H}_{4+2n}(\mu)$ for the single inclusive parton production obeys the same DGLAP evolution as $  [\vec{{\cal D}}_{J_L = -6 - 2n }^{\text{DGLAP}}]_R $ in Eq.~(\ref{eq:Drun}), since the moment variable is given by $J-1 = -J_L-2 = 4+2n$. By consistent condition $d\text{EEC}/d\ln\mu^2 = 0$, the jet function must then evolve as 
\bea 
\frac{d\vec{{\cal J}}^{(n)}_{1,1}(\mu^2;\Lambda_{\rm QCD}^2)}{d\ln\mu^2} 
= 
\vec{{\cal J}}^{(n)}_{1,1}(\mu^2;\Lambda_{\rm QCD}^2)\cdot \widehat{\gamma}_{5+ 2n}  \,,\quad
\eea 
where $\widehat{\gamma}_J$ is the DGLAP anomalous dimension matrix. This evolution correctly matches the known evolution of the DFF at non-zero angular separation~\cite{deFlorian:2003cg}. 
This agreement in the scaling behavior from both the DFF and OPE perspectives validates the connection between these two independent approaches. 

The relation is straightforwardly generalized to the energy correlators with higher energy weights in Eq.~(\ref{eq:geec_LR_OPE}), which will relate to higher moments of the DFF, i.e., ${\cal J}^{(n)}_{\alpha,\beta}=\vec{R}^{\alpha,\beta}_n$.

\section{Phenomenological Studies} %

In this section, we validate our theoretical setup by comparing the LL prediction from Eq.~\eqref{eq:resum_result} with Monte Carlo simulations from {\tt Pythia} 8.2~\cite{Sjostrand:2014zea}. 
We begin by studying global energy correlators in $e^+e^-$ annihilation, using $e^+e^-\to \gamma/Z^\ast \to X(q\bar{q})$ and $e^+e^- \to h^\ast \to X(gg)$ to source the quark and gluon initiated energy correctors, respectively. 
 \begin{figure*}[htbp]
  \centering  
  \begin{subfigure}{0.48\textwidth}  
    \centering
    \includegraphics[width=\linewidth]{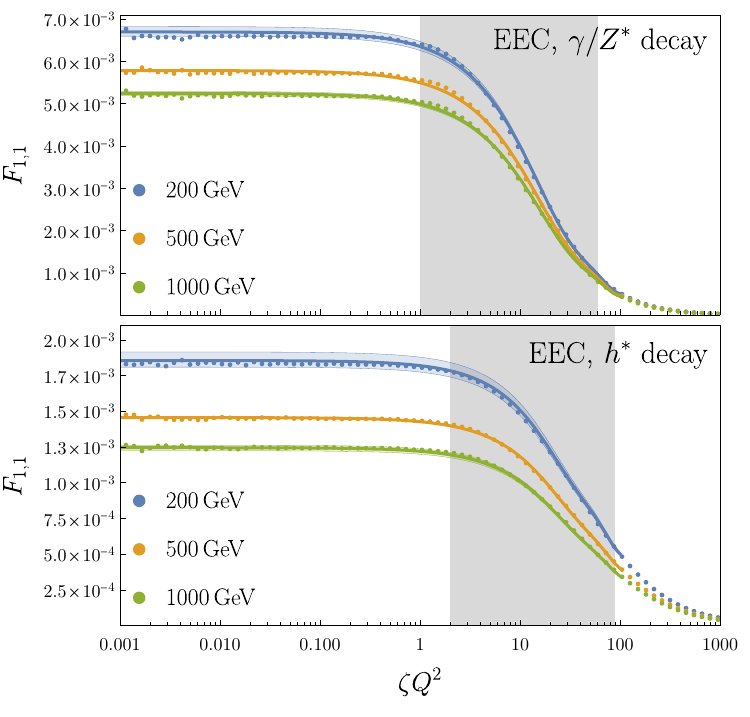}  
    \label{fq:eec_plots}
  \end{subfigure}  
    \hfill    
  \begin{subfigure}{0.48\textwidth}  
    \centering
    \includegraphics[width=\linewidth]{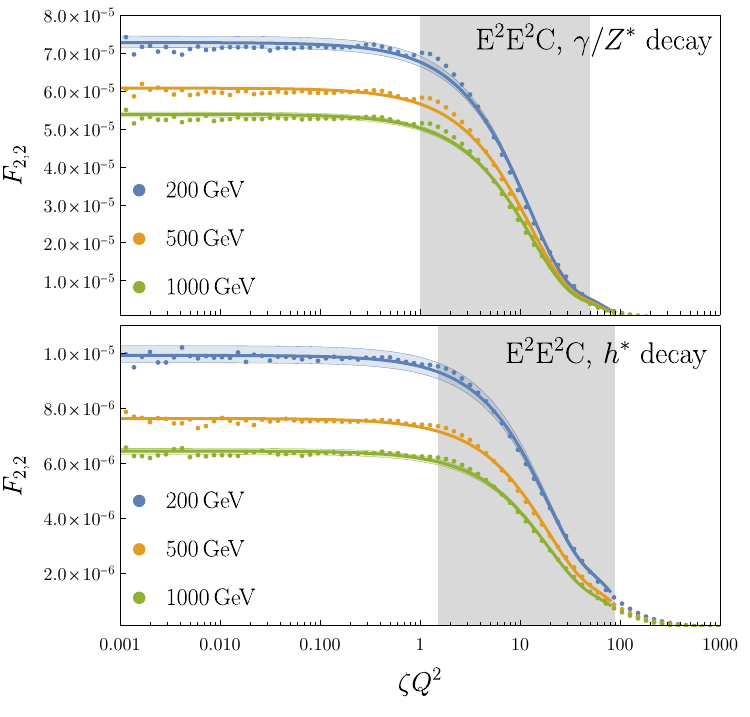}  
    \label{fg:e2e2c_plots}
  \end{subfigure}

  \caption{ Validation of the LL-evolved light-ray OPE against {\tt Pythia} simulations in $e^+e^-$ annihilation. 
        The left column shows the $F_{1,1}$ correlator, and the right column shows the $F_{2,2}$ correlator for both quark-initiated (top) and gluon-initiated (bottom) jets. 
        The solid lines are theoretical predictions obtained by fitting the OPE in Eq.~\eqref{eq:geec_LR_OPE} to the {\tt Pythia} data (dots) at $Q = 500\>{\rm GeV}$ and then evolving the result to $Q=200$ and $1000\>{\rm GeV}$ with a scale choice of $\mu_H = Q/4$. The bands are scale uncertainties obtained by varying $\mu$ up and down by a factor of $2$.}  
  \label{fg:eec_ee}
\end{figure*}

Fig.~\ref{fg:eec_ee} presents this comparison for the $F_{1,1}$ (left column) and $F_{2,2}$ (right column) correlators at the center-of-mass energies of $Q = 200\>{\rm GeV}$, $500\>{\rm GeV}$ and $1000\>{\rm GeV}$. As detailed in the figure caption, our theoretical predictions are derived by fitting the light-ray OPE to the $Q = 500\>\rm{GeV}$ simulation data and then evolving the result to the other energies. Details on the initial condition fitting are given in the SM.
Across all scenarios, the predictions (solid lines) show excellent agreement with the Monte Carlo simulation (dots). It is interesting to note that the convergence of the light-ray OPE fitting breaks down around $Q^2\zeta \sim {\cal O}(50)\>{\rm GeV^2}$, setting the boundary between the post- and pre-confinement regimes.

Our framework automatically extends to 
energy correlators measured inside jets at the LHC, as well as
track-based measurements~\cite{Chang:2013rca,Li:2021zcf}, which offer superior angular resolution~\cite{Bossi:2025xsi}, making them more suitable for investigating the post-confinement regime. Detailed studies on this, along with a discussion on its potential application to the precise determination of $\alpha_s$, can be found in the SM.

\section{Conclusion and Outlook}

We developed a light-ray OPE approach to studying the $Q^2$-scaling behavior of the EEC at small angles, focusing on the post-confinement and dynamical transition regimes. We successfully described the $Q^2$-scaling observed in Pythia simulations for a range of energies and collision processes in the transition and the post-confinement plateau, where in the latter case, the quantum scaling is determined by the DGLAP anomalous dimension with $J=5$. We also discuss energy correlators with more general weights, E$^{\alpha}$E$^{\beta}$C. 

Furthermore, we established a direct connection between the light-ray OPE formalism and the mass-dependent dihadron fragmentation function framework, demonstrating that the OPE coefficients correspond to moments of (the derivative of) the DFF.

Our work opens up several avenues for future exploration. Perhaps the primary priority is to test the quantum scaling behavior predicted in this paper experimentally. This would likely require track-based measurement to reach high accuracy in the post-confinement regime, for which our predictions remain valid. We further propose to extract $\alpha_s$ from precise scaling measurements. Our findings also suggest that the nearside EEC provides a direct probe of the mass-dependent DFF. This offers a novel experimental method to measure these nonperturbative distribution functions, which are crucial for understanding the hadronization process. The established link between the phenomenological DFF formalism and the theoretical light-ray OPE creates a new paradigm for studying nonperturbative QCD. %

{\it Note added.}— While this article was being completed, we became aware of Ref.~\cite{Lee:2025okn}, where the scaling behavior in the near-side region was also studied using the transverse momentum dependent dihadron fragmentation formalism.

\begin{acknowledgments}

\textbf{\textit{Acknowledgement.}} 
 This work was initiated during a C$^3$NT program on ``New Opportunities in Particle and Nuclear Physics with Energy Correlators" at CCNU, Wuhan, May 7-16, 2025. We thank Wei-Yao Ke and Xin-Nian Wang for the warm hospitality. We also thank Petr Kravchuk for related discussions. HC is supported by the U.S. Department of Energy, Office of Science, Office of Nuclear Physics under grant Contract Number DE-SC0011090.
FY is supported by the Office of Science of the U.S. Department of Energy under Contract No. DE-AC02-05CH11231.  
HXZ is supported by the National Natural Science Foundation of China under contract
No. 12425505 and The Fundamental Research Funds for the Central Universities, Peking University. X.L. is supported by the National Natural Science Foundation of China under Contract No. 12175016 and the Fundamental Research Funds for the Central Universities, Beijing Normal University. DSD is supported in part by Simons Foundation grant 488657 (Simons Collaboration on the Nonperturbative Bootstrap) and the U.S. Department of Energy, Office of Science, Office of High Energy Physics, under Award Number DE-SC0011632. CHC is supported by a Kadanoff fellowship at the University of Chicago.

 \end{acknowledgments}

\bibliographystyle{h-physrev}   
\bibliography{refs}

\clearpage
\appendix

\begin{widetext}
\section{Supplemental Materials}  

\subsection{Details of All-order Factorization Formula}

The discussions from both light-ray OPE and dihadron fragmentation perspectives in this Letter can be summarized as the following factorization formula in the collinear limit, covering both post-confinement, pre-confinement and transition regions
\begin{align}\label{eq: all-order-factorization}
    \GEEC{\alpha}{\beta}(\zeta,Q) = \frac{1}{\zeta}\int_0^1 dx\, x^{\alpha+\beta}  \vec{J}_{\alpha,\beta}(\frac{\zeta x^2 Q^2}{\Lambda_{\text{QCD}}^2},\frac{\Lambda_{\text{QCD}}^2}{\mu^2};\mu^2)\cdot \vec{H}(x,\frac{Q^2}{\mu^2};\mu^2) + \cdots \,,
\end{align}
where $\vec{J}_{\alpha,\beta}$ is the non-perturbative jet function. Here, we consider massless QCD and assume $\Lambda_{\text{QCD}}$ is the only characteristic scale. The functional dependence of $\vec{J}_{\alpha,\beta}$ is constrained by Lorentz symmetry -- applying a collinear boost to the jet function, the transformations of the small opening angle $\sim \sqrt{\zeta}$ and jet energy $E_J\sim x Q$ have the opposite scaling
\begin{align}
\sqrt{\zeta} \to \lambda^{-1} \sqrt{\zeta}, \qquad E_J \to \lambda E_J\,.
\end{align}
After factorizing out classical scaling dependence, $\zeta x^2 Q^2$ is the only boost invariant combination of kinematic variables in jet function.
This is also the underlying symmetry origin of the small jet radius factorization formula in~\cite{Lee:2024tzc}.

Eq. \eqref{eq: all-order-factorization} is the generalization of collinear factorization of EEC in perturbation theory~\cite{Dixon:2019uzg} and the same form as the proposal of $N$-point projective energy correlator factorization in the pre-confinement region~\cite{Chen:2024nyc}. Taking $N$-point projective energy correlator jet function $\vec{J}_N$ as an example, the dominant terms in the expansion of its jet function $\vec{J}_N$ in the $\Lambda_{\text{QCD}}^2\ll \zeta Q^2$ limit are
\begin{align}
\vec{J}_N(\frac{\zeta x^2 Q^2}{\mu^2},\frac{\Lambda_{\text{QCD}}^2}{\mu^2};\mu^2) = \vec{J}^{\text{P.T.}}_N(\frac{\zeta x^2 Q^2}{\mu^2};\alpha_s(\mu^2)) + \frac{\Lambda_{\text{QCD}}}{\sqrt{\zeta} x Q} \vec{J}^{(1)}_N(\frac{\zeta x^2 Q^2}{\mu^2}, \frac{\Lambda_{\text{QCD}}^2}{\zeta x^2 Q^2};\mu^2)+\cdots\,,
\end{align} 
where the leading term $\vec{J}^{\text{P.T.}}$ is perturbatively calculable~\cite{Chen:2020vvp,Chen:2023zlx} and leading hadronization correction $\vec{J}^{(1)}$ is discussed in~\cite{Chen:2024nyc,Lee:2024esz}. Note that $\vec{J}^{(1)}$ may have mild logarithmic dependence on $\frac{\Lambda_{\text{QCD}}^2}{\zeta x^2 Q^2}$ due to hadronization dynamics.
$\GEEC{\alpha}{\beta}$ in the pre-confinement region is more complicated due to its collinear unsafety and  can be found in the upcoming work~\cite{GLMUpcoming}.

In this Letter, we study the expansion in the post-confinement region $\zeta Q^2 \ll \Lambda_{\text{QCD}}^2$. To establish a connection with \eqref{eq:geec_LR_OPE} in the absence of QED radiation and hadron decay (see a later section on QED corrections for relevant discussion), we can expand $\vec{J}_{\alpha,\beta}$ as
\begin{align}
\vec{J}_{\alpha,\beta}(\frac{\zeta Q^2 }{\Lambda_{\text{QCD}}^2},\frac{\Lambda_{\text{QCD}}^2}{\mu^2};\mu^2)= \sum_{n\geq 0} (\zeta Q^2 )^{n+1} \vec{\cal J}_{\alpha,\beta}^{(n)}= \sum_{n\geq 0} (\zeta Q^2 )^{n+1} \vec{R}^{\alpha,\beta}_n (\mu^2;\zeta Q^2, \Lambda_{\text{QCD}}^2)\,.
\end{align}
Here, the $\zeta Q^2$ dependence in $\vec{R}^{\alpha,\beta}_n$ manifests as the ratio $\frac{\zeta Q^2}{\Lambda_{\text{QCD}}^2}$ and is expected to be logarithmic (and calculable if we know the corresponding hadron EFT). In the language of light-ray OPE, the logarithmic $\zeta$-dependence requires including derivatives of perturbative detectors $\partial_{J_L}^k \vec{\cD}^\DGLAP_{J_L}$, see \cite{Chen:2023zzh} for related discussion. In this Letter, we keep these contributions implicit for simplicity, as well as due to the weak interactions between hadrons in the post-confinement region.

\subsection{Details of Series Approximation and Scaling Prediction}

In this section, we provide more details on the initial condition fitting in Fig.~\ref{fg:eec_ee}. %
The generalization to the jet case in $pp$ collisions is straightforward.
Given the light-ray OPE formalism in Eq.~\eqref{eq:geec_LR_OPE}, we adopt the following series ansatz to approximate the shapes $F_{\alpha,\beta}$  %
\begin{align}\label{eq:series_ansatz}
    F_{\alpha,\beta}(\zeta Q^2, Q^2) \approx \sum_{n=0}^{N} c_{\alpha,\beta}^{(n)}(Q)\times (\zeta Q^2)^n\,.
\end{align}

In Fig.~\ref{fg:eec_ee}, %
we choose $Q=500\>{\rm GeV}$ to be the initial condition, and set $N=8$ to fit the region up to $\zeta Q^2 \lesssim 100\, \mathrm{GeV}^2$. As a comparison, Fig.~\ref{fg:fit_comparison} shows a fit with $N=2$. In both cases, the coefficients $c_{\alpha,\beta}^{(n)}$ are found to decay quickly as we go to higher powers of $\zeta Q^2$. 

Typically, with more terms included, the ansatz provides a better approximation at large values of $\zeta Q^2$. However, we note that the convergence of this expansion breaks down as we approach the perturbative pre-confinement regime. Therefore, the convergent radius of the series estimates the boundary of the transition.

\begin{figure*}[htbp]
  \centering  
  \begin{subfigure}{0.48\textwidth}  
    \centering
    \includegraphics[width=\linewidth]{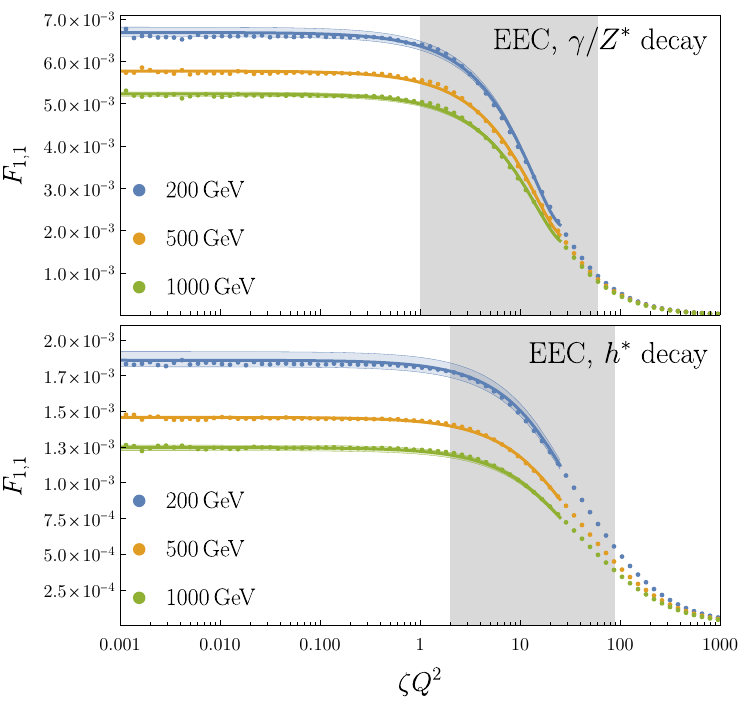}  
    \label{fg:eec_3_param}
  \end{subfigure}  
    \hfill    
  \begin{subfigure}{0.48\textwidth}  
    \centering
    \includegraphics[width=\linewidth]{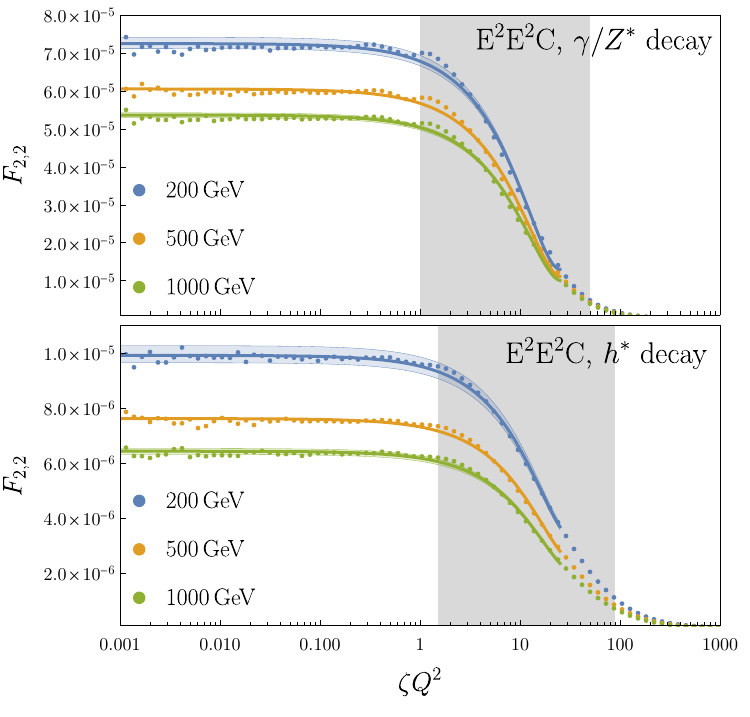}  
    \label{fg:e2e2c_3_param}
  \end{subfigure}
  \caption{Fitting initial condition at $Q=500\>{\rm GeV}$ with $N=2$. Other curves for $Q=200\>{\rm GeV}$ and $1000\>{\rm GeV}$ are obtained by evolution. }  
  \label{fg:fit_comparison}
\end{figure*}

Once we obtain $c_{\alpha,\beta}^{(n)}$ at one center-of-mass energy $Q_0$, we can predict the shape of $F_{\alpha,\beta}$ at generic energy $Q$. This prediction greatly simplifies at leading logarithmic accuracy --- the evolution kernel \eqref{eq:LL_kernel} factorizes, which converts \eqref{eq:resum_result} to the form
\begin{align}\label{eq:resum_LL_form}
    F_{\alpha,\beta}(\zeta Q^2, Q^2) =\sum_{n\geq 0} (\zeta Q^2)^n \vec{V}^{\alpha,\beta}_n \cdot  \left[\alpha_s(\mu_H^2)\right]^{\widehat{\gamma}^{(0)}_{-J_L-1}/\beta_0} \cdot \frac{\langle  [\vec{\mathcal{D}}^{\text{DGLAP}}_{J_L}]_R(\mu_H^2)\rangle_Q}{Q^{-J_L-2}}\Big|_{J_L=-\alpha-\beta-4-2n},
\end{align}
where $\vec{V}_n^{\alpha,\beta}=\vec{V}_n^{\alpha,\beta}(\Lambda_{\text{QCD}}^2) = \vec{R}_n^{\alpha,\beta}(\mu_J^2,\Lambda_{\text{QCD}}^2)\cdot \left[\alpha_s(\mu_J^2)\right]^{-\widehat{\gamma}^{(0)}_{-J_L-1}/\beta_0}$ is independent of $\mu_J$ due to RG invariance. To manifest hard scale choice $\mu_H \sim \mathcal{O}(Q)$, we use the parametrization $\mu_H = \kappa Q$. Compared with the series ansatz \eqref{eq:series_ansatz}, we have the relation
\begin{align}
c^{(n)}_{\alpha,\beta}(Q) = \vec{V}^{\alpha,\beta}_n \cdot  \left[\alpha_s(\kappa^2 Q^2)\right]^{\widehat{\gamma}^{(0)}_{-J_L-1}/\beta_0} \cdot \frac{\langle  [\vec{\mathcal{D}}^{\text{DGLAP}}_{J_L}]_R(\kappa^2 Q^2)\rangle_Q}{Q^{-J_L-2}}\Big|_{J_L=-\alpha-\beta-4-2n}\,.
\end{align}

For $\gamma/Z^*\to q\bar{q}$ and $h^*\to gg$ processes, we use the leading-order matrix elements
\begin{align}
    \frac{\langle  [\vec{\mathcal{D}}^{\text{DGLAP}}_{J_L}]_R(\mu_H^2)\rangle^{\gamma/Z^*}_Q}{Q^{-J_L-2}} = 2^{2+J_L} \begin{pmatrix}
        1\\ 0
    \end{pmatrix}\,,\qquad 
    \frac{\langle  [\vec{\mathcal{D}}^{\text{DGLAP}}_{J_L}]_R(\mu_H^2)\rangle^{h^*}_Q}{Q^{-J_L-2}} = 2^{2+J_L} \begin{pmatrix}
        0\\ 1
    \end{pmatrix}\,.
\end{align}
Using $c^{(n)}_{\alpha,\beta}$ extracted from both $\gamma/Z^*\to q\bar{q}$ and $h^*\to gg$ at $Q=Q_0$, we can solve the linear equations to extract $\vec{V}^{\alpha,\beta}_n$
\begin{align}
   & \left[c^{(n)}_{\alpha,\beta}(Q_0)\right]_{\gamma/Z^*} = \vec{V}^{\alpha,\beta}_n \cdot  \left[\alpha_s(\kappa^2 Q_0^2)\right]^{\widehat{\gamma}^{(0)}_{-J_L-1}/\beta_0} \cdot \begin{pmatrix}
        2^{J_L+2}\\ 0
    \end{pmatrix}\Big|_{J_L=-\alpha-\beta-4-2n}\, \\
    & \left[c^{(n)}_{\alpha,\beta}(Q_0)\right]_{h^*} = \vec{V}^{\alpha,\beta}_n \cdot  \left[\alpha_s(\kappa^2 Q_0^2)\right]^{\widehat{\gamma}^{(0)}_{-J_L-1}/\beta_0} \cdot \begin{pmatrix}
        0 \\ 2^{J_L+2}
    \end{pmatrix}\Big|_{J_L=-\alpha-\beta-4-2n}\,.
\end{align}
Then we can use \eqref{eq:resum_LL_form} to make predictions at various $Q$ and find a good agreement with {\tt Pythia} simulation results. Though we only present leading logarithmic resummation in this Letter, systematically extending prediction at higher logarithmic accuracy is readily obtained by currently known results in the literature.

\subsection{Predictions for Energy Correlators inside Jets at the LHC}
The theoretical setup can be similarly applied to energy correlators measured inside jets produced in $pp$ collisions. For this study, we simulate $pp\to Z q$ and $pp\to Z g$ events for the quark and gluon jets, respectively. The energy correlators are measured within anti-${k_t}$ jets clustered with a radius parameter $R = 0.6$. We impose an almost back-to-back topology by requiring the two leading jets to have an azimuthal separation $\Delta\phi_{j_1,j_2} > 2$. The intra-jet correlator is defined as 
\begin{equation}
F_{\alpha,\beta}(R_L, R, E_J) = \sum_{(a,b)}
\int \frac{d\sigma_{J,ab}}{\sigma_{J}} \frac{E_a^\alpha E_b^\beta}{E_J^{\alpha+\beta-2}}
\delta(R_L^2 - \hat{R}_{ab})
\end{equation}
where the sum runs over all hadrons $a,b$ within the leading two jets.   
Here, $E_J$ is the jet energy and $\hat{R}_{ab}$ is the operator that gives the angular separation between the measured hadrons $(\Delta\eta_{ab})^2 + (\Delta\phi_{ab})^2$. 

 \begin{figure*}[htbp]
  \centering  
  \begin{subfigure}{0.48\textwidth}  
    \centering
    \includegraphics[width=\linewidth]{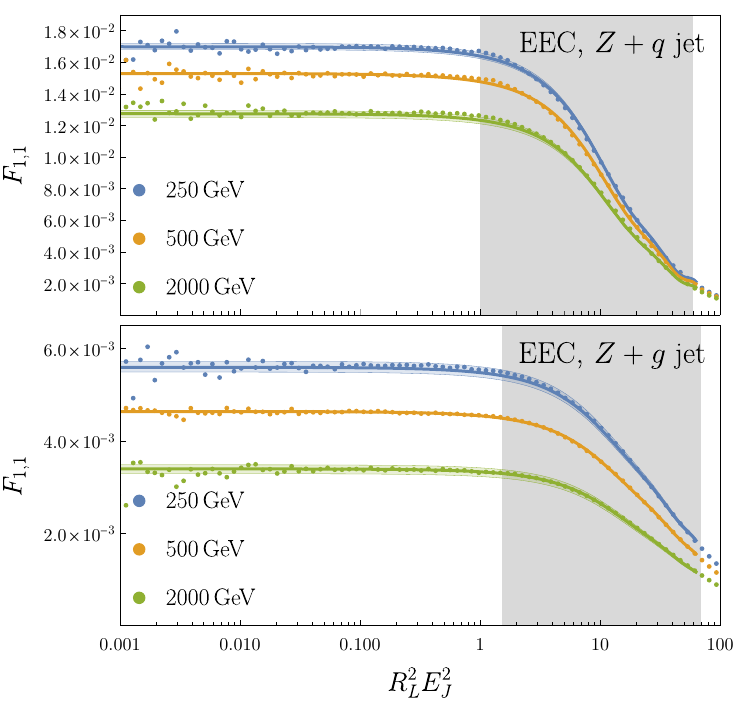}  
    \label{fq:e2e2c}
  \end{subfigure}  
    \hfill    
  \begin{subfigure}{0.48\textwidth}  
    \centering
    \includegraphics[width=\linewidth]{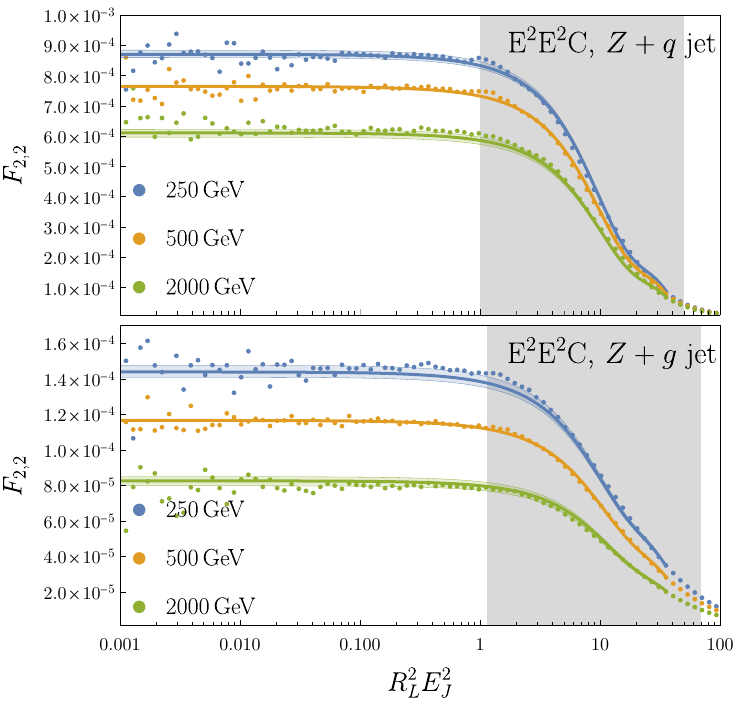}  
    \label{fg:e2e2c}
  \end{subfigure}

  \caption{Validation of the LL-evolved light-ray OPE against {\tt Pythia} simulations in $pp$ collisions. The plots show $F_{1,1}$ (left) and $F_{2,2}$ (right) for quark jets (top) and gluon jets (bottom). Predictions (solid lines) are derived by fitting to the $E_J = 500\>{\rm GeV}$ data (dots) and evolving to other values of $E_J$.}  
  \label{fg:eec_lhc}
\end{figure*}
As shown in Fig.~\ref{fg:eec_lhc}, we follow the same strategy used in the $e^+e^-$ analysis. We fit the light-ray OPE expansion in Eq.~\eqref{eq:series_ansatz} to the Monte Carlo data with $N=8$ for jets with $E_J = 500\>{\rm GeV}$, and predict the correlators for other jet $E_J$ bins through LL evolution, using a scale choice $\mu = E_J R$. For both $F_{1,1}$ and $F_{2,2}$, we again observed remarkable agreement between our theoretical predictions and the {\tt Pythia} simulation.

\subsection{Predictions for Track Data}
The resummation formalism presented in Eq.~\eqref{eq:resum_result} seamlessly extends to track-based measurements, which exclusively utilize charged particles for EEC determination~\cite{Chen:2020vvp,Li:2021zcf,Jaarsma:2023ell}. Fig.~\ref{fg:eec_height_running} provides a compelling validation of the LL evolution against Pythia track data, clearly illustrating the quantum scaling of the post-confinement plateau across varying $Q$ values. 
 \begin{figure*}[htbp]
  \centering  
  \begin{subfigure}{0.48\textwidth}  
    \centering
    \includegraphics[width=\linewidth]{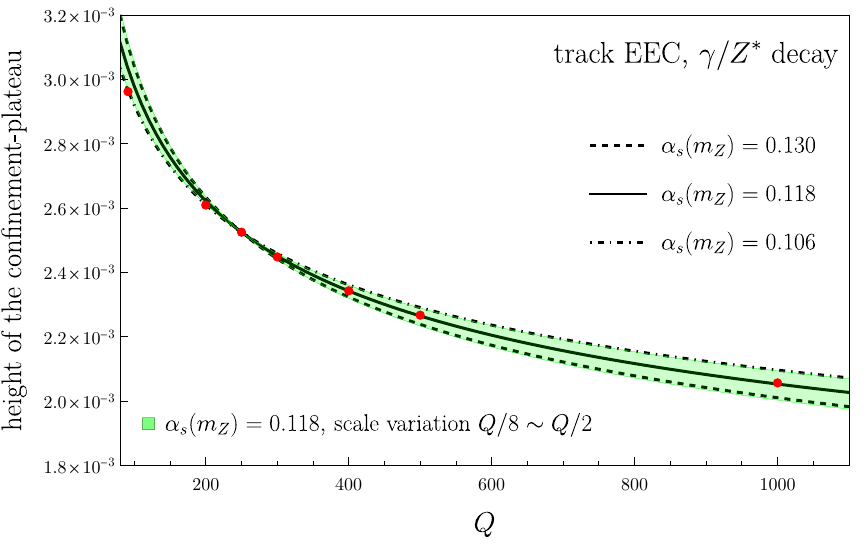}  
    \label{fg:eec_q_height_running}
  \end{subfigure}  
    \hfill    
  \begin{subfigure}{0.48\textwidth}  
    \centering
    \includegraphics[width=\linewidth]{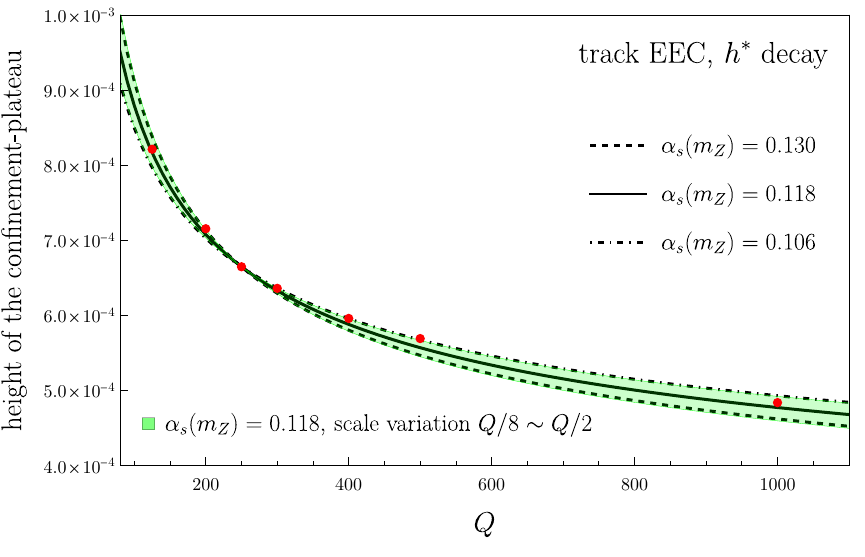}  
    \label{fg:eec_g_height_running}
  \end{subfigure}

  \caption{Quantum scaling of the post-confinement plateau for EEC using tracks. }  
  \label{fg:eec_height_running}
\end{figure*}
For this evolution, the plateau height at $Q=250\>{\rm GeV}$ was used as the initial condition and subsequently evolved to other $Q$ values using Eq.~\eqref{eq:resum_result} with only the $n = 0$ term. We observed excellent agreement with the Monte Carlo data, represented by the red dots. %

This scaling behavior offers an immediate and significant application for the precise determination of $\alpha_s$, given that initial values of $\alpha_s$ substantially influence the shape of the scaling across different Q values. As a proof of concept, Fig.~\ref{fg:eec_height_running} illustrates this by showing the evolution with $\alpha_s(m_Z)$ varied by $\pm10/\%$, in dashed and dotted-dashed lines, respectively. We observe that this variation in $\alpha_s$ results in a shape change comparable to the LL error band. We thus anticipate that, with advancements in theoretical precision, this will provide a novel and precise method for measuring $\alpha_s$ at LEP or LHC.

\subsection{Including QED Corrections}

While our previous Monte Carlo studies neglected QED effects to isolate the QCD dynamics of the EEC, such interactions are crucial for a complete physical description. Processes like neutral pion decay $\pi^0 \to \gamma \gamma$ and QED final-state radiation introduce unique features into the EEC spectrum. These photon-induced signatures, which are typically masked by limited angular resolution, could be identified with future analysis. Fig.~\ref{fg:eec_QED} demonstrates this by presenting the EEC distribution from a {\tt Pythia} simulation where these QED effects have been enabled.

The leading QED corrections in the collinear limit come from the processes in which a hadron emits an almost collinear photon and both the hadron and the photon are detected by the two energy detectors. This process has a $1/\theta^2$ collinear divergence, where $\theta$ is the angle between the hadron and the emitted photon. This divergence allows us to predict the Lorentz spin of the corresponding detector in the OPE,
\begin{align}\label{eq:LR_OPE_QED}
\cE(\boldsymbol{n}_1)\cE(\boldsymbol{n}_2) = \alpha_e\frac{A_{\text{QED}}}{\theta^2}\cD^{\text{DGLAP}_h}_{-4}(\boldsymbol{n}_2) + \ldots.
\end{align}
Here, $\cD^{\text{DGLAP}_h}_{J_L}$ is a DGLAP hadron detector that measures hadrons weighted by $E^{-2-J_L}$. The Lorentz spin $J_L=-4$ is fixed by the $\frac{1}{\theta^2}$ factor. The fine structure constant $\alpha_e$ shows that the process is due to QED effects. $A_{\text{QED}}$ is the OPE coefficient, and $\ldots$ denotes terms that are subleading in the collinear limit $\boldsymbol{n}_1 \to \boldsymbol{n}_2$. The coefficient $A_{\text{QED}}$ can be computed using the method in \cite{Chen:2021gdk}, but we will not need the explicit expression for the discussion here.

As argued in the main text, the operator in Eq.~\eqref{eq:LR_OPE_QED} can be matched onto detectors in perturbative QCD, and the leading term should be the twist-2 DGLAP detectors with $J_L=-4$. As a result, the effect of this term in the post-confinement region would be a $1/\theta^2$ term whose $Q^2$-scaling is governed by the DGLAP anomalous dimension at $J=3$. As shown in Fig.~\ref{fg:eec_QED}, this prediction agrees with the {\tt Pythia} simulation at small angles, correctly reproducing both the singular $1/\theta^2$ behavior and the $Q^2$-scaling.

Additional hadron interactions could lead to other types of detectors in the light-ray OPE in the EFT of hadron. For example, the neutral pion decay $\pi^0 \to \gamma\gamma$ gives rise to a term of the form $\frac{\alpha_e^2}{f_\pi^2 m_\pi^2}\theta^4 \cD^{\text{DGLAP}_h}_{-10}$. However, this term has the same $\theta$-dependence and $J_L$ as the double-twist operator $[\cE \cE]_{n,0}$ with $n=2$, and thus would not affect the fit. It would be interesting to explore other hadron interactions that further modify the ansatz used in the fit.

Compared with Fig.~\ref{fg:eec_ee}, we notice that QED radiation and hadron decay non-trivially modify the shape of EEC. The Laurent series approximation is not the optimal ansatz for non-perturbative functions. In other cases, we can use the equivalent all-order factorization formula
\eqref{eq: all-order-factorization}.

 \begin{figure*}[htbp]
  \centering  
  \begin{subfigure}{0.48\textwidth}  
    \centering
    \includegraphics[width=\linewidth]{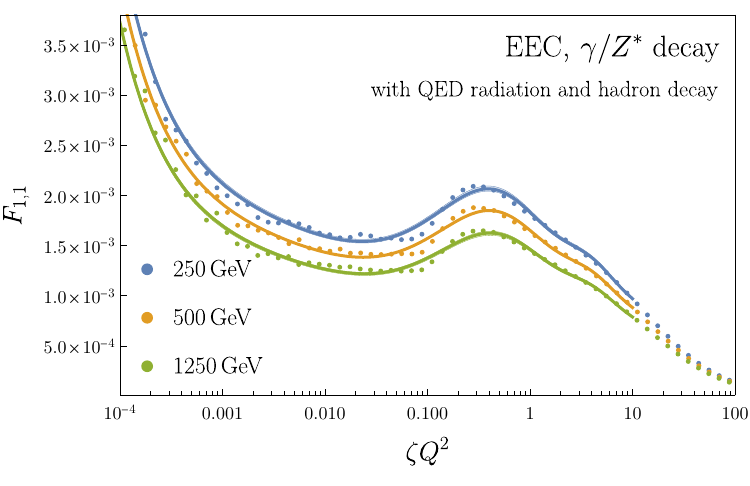}  
    \label{fg:eec_q_QED}
  \end{subfigure}  
    \hfill    
  \begin{subfigure}{0.48\textwidth}  
    \centering
    \includegraphics[width=\linewidth]{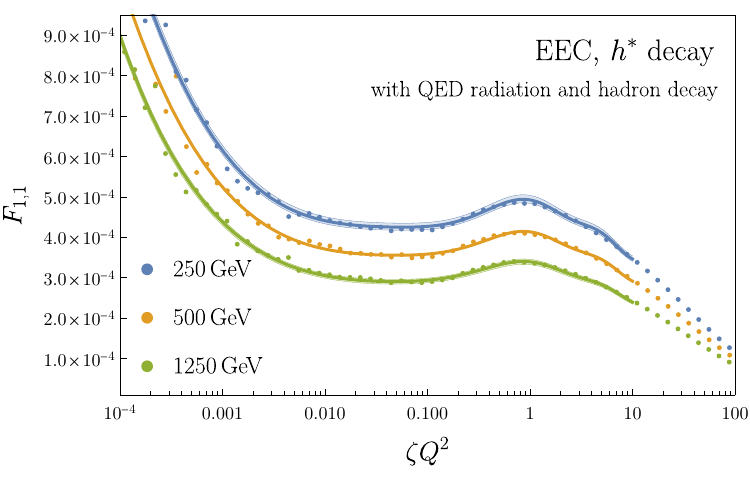}  
    \label{fg:eec_g_QED}
  \end{subfigure}

  \caption{Validation of the LL-evolved light-ray OPE against {\tt Pythia} simulations in $e^+e^-$ annihilation, with QED radiation and hadron decay turned on. Due to the presence of QED radiation and the pion-decay bump, the Laurent series ansatz requires a significantly larger number of terms to approximate this curve. Here, we employ 24 terms and demonstrate the evolution for $\zeta Q^2 < 10 \,\mathrm{GeV}^2$. }  
  \label{fg:eec_QED}
\end{figure*}

\end{widetext}

\end{document}